\documentclass{article}
\usepackage{graphicx}
\usepackage{epstopdf}
\usepackage{amsmath}
\usepackage{amssymb}
\usepackage{mathptmx}
\begin{document}

\vspace*{3cm} \thispagestyle{empty}
\vspace{5mm}

\noindent \textbf{\Large A statistical mechanical problem in Schwarzschild spacetime}\\

\noindent \textbf{\normalsize Peter Collas}\footnote{Department of Physics and Astronomy, California
State University, Northridge, Northridge, CA 91330-8268. Email: peter.collas@csun.edu.}
\textbf{\normalsize and David Klein}\footnote{Department of Mathematics, California State University,
Northridge, Northridge, CA 91330-8313. Email: david.klein@csun.edu.}\\

\vspace{4mm} \parbox{11cm}{\noindent{\textbf{Abstract} We use Fermi coordinates to calculate the canonical partition function for an ideal gas in a circular geodesic orbit in Schwarzschild spacetime.  To test the validity of the results we prove theorems for limiting cases.  We recover the Newtonian gas law subject only to tidal forces in the Newtonian limit.  Additionally we recover the special relativistic gas law as the radius of the orbit increases to infinity. We also discuss how the method can be extended to the non ideal gas case.}\vspace{5mm}

}

\setlength{\textwidth}{27pc}
\setlength{\textheight}{43pc}

\section*{{\normalsize 1 Introduction}}
Contributions to the development of a theory of relativistic statistical mechanics, including attempts to find a fully covariant theory, date back to the early part of the 20th century.  A useful summary of contributions up to 1967 is given by Hakim \cite{H67}. Since that time
distinct proposals for a general framework for  relativistic statistical mechanics have continued to
emerge.  For example, Horwitz, Schieve and Piron \cite{HSP81} constructed classical and quantum Gibbs ensembles in the context of special relativity using ``historical time'' and a statistics of events
off the mass shell. Miller and Karsch \cite{MK81} applied the constraint formalism to many-particle
special relativistic systems in statistical mechanics in order to include two-particle
interactions.  Droz-Vincent \cite{D-V97} considered a statistics of orbits consistent with Hakim's approach
including a tentative definition of equilibrium.  More recently, Montesinos and Rovelli \cite{MR01} proposed a theory of statistical mechanics that is not necessarily tied to the notion of energy or a preferred time axis and which is consequently not necessarily governed by the notion of temperature.

In this paper, we consider an ideal gas enclosed in a container in orbit around the central mass in Schwarzschild spacetime.  It is assumed that the gas particles, along with the container, do not significantly alter the background Schwarzschild metric.We develop a statistical mechanical formalism that yields approximations of the canonical Helmholtz free energy for the volume of ideal gas. These approximations involve cross terms between space coordinates, momentum coordinates, and Riemann curvature tensor components, but the expressions are still computationally accessible enough to allow comparision to their special relativistic and Newtonian analogs. We test the validity of these expressions by proving rigorous results for special relativistic and Newtonian limits. Technical issues related to the existence of the thermodynamic limit are discussed in Remark 1 below.

Our method begins with the calculation of the Schwarzschild metric in Fermi coordinates along the circular geodesic orbit.  For that purpose, we use the results of \cite{MM63} and \cite{PP82}.  The choice of a circular orbit in the  spherically symmetric Schwarzschild spacetime makes it plausible on physical grounds that the gas  will achieve equilibrium. Since Fermi coordinates are Minkowskian to first order near the orbit, the statistical mechanical partition function should closely resemble the analogous
well-known  expression in special relativity (see e.g. \cite{Pauli}), but with  corrections
to second  order.   We note  that the term  ``ideal gas'' is somewhat misleading in the context of
general  relativity.  This is because a volume of gas subject to no forces is still affected by the
curvature  of spacetime, and this corresponds to a Newtonian gas subject to tidal forces, but
otherwise ``ideal.''  We elaborate on this correspondence in Sec. 4.

Our formulation and results may be compared with  \cite{MR01} and \cite{C64}.  In Section 5 of \cite{MR01} it was 
shown that even for the ideal gas in Minkowski spacetime, ``there is no Lorentz-invariant thermal state: a 
thermal state is in equilibrium in a \textit{preferred} Lorentz frame of reference, and therefore breaks 
Lorentz invariance.''  It was further argued in \cite{MR01} that when the gas is enclosed in a box, the coordinate system in which the center of mass of the system is at rest is the preferred Lorentz frame, and the canonical distribution density for an ideal gas in Minkowski
spacetime then reduces to a product of terms of the form $\exp \beta p_{0}$ (where $p_{0}$ is the
negative of the energy of a particle and $\beta$ is the inverse temperature).  Analogously, we work in the locally Minkowskian Fermi normal coordinate system of the rest frame of the box of gas in Schwarzschild spacetime. We expect this to be the preferred reference frame for equilibrium. In Section 8 of \cite{C64}, Chernikov gave a solution to the kinetic Boltzmann equation for the single-particle equilibrium distribution of an ideal gas in the Schwarzschild coordinates. In contrast to the methods of \cite{C64}, we use statistical mechanical arguments for derivation of the canonical partition function.   We make further comparisons in Section 5.  The time-dependence of the Fermi metric considered here necessarily introduces complications.

In Sec. 2 we introduce notation and compute the Schwarzschild metric in Fermi coordinates to 
second order.  In Sec. 3 we derive the volume form for phase space, give a theorem for coordinate
independence, and establish a local version of Liouville's Theorem on a $6N$-dimensional phase
space for $N$ particles. We then find approximations for the partition function and free energy for the ideal  gas.  The proofs of the main results of this section appear in Appendices A and B.  In Sec. 4 we prove that the non relativistic Newtonian limit of our general relativistic free energy
function is the expected result corresponding to classical tidal forces. We also prove that the limit of our general relativistic estimate of the free energy, as the radius of the orbit goes to
infinity, equals the free energy for the special relativistic case found by Pauli \cite{Pauli}.  Sec. 5 is
devoted to conclusions.

\section*{{\normalsize 2 Fermi coordinates}}
Throughout we use the sign conventions of Misner, Thorne and Wheeler \cite{Misner} and express the
Schwarzschild line element in terms of the usual variables as,
\begin{equation}
ds^{2}=-\left(1-\frac{2M}{r}\right)dt^{2}+\displaystyle\frac{dr^{2}}{\displaystyle
\left(1-\frac{2M}{r}\right)}+r^{2}(d\theta^{2}+\sin^{2}\theta d\phi^{2}).
\label{a1}
\end{equation}

Consider a container of gas molecules {in orbit  around the central mass $M$ in Schwar-zschild spacetime.  We assume that neither the gas nor the container significantly contribute to the gravitational field, so that the Schwarzschild metric is effectively unaltered by their presence.  We assume further that a center of mass of the container of the gas may be identified, and that it follows (or very nearly follows) a circular geodesic orbit $\mathcal {C}$ of radius $r>3M$\footnote{Since circular orbits are stable only for $r>6M$, it may be argued that equilibrium is possible only for such orbits.} with $\theta=\pi/2$; (a discussion of the existence of a center of mass related to our circumstances is given in \cite{ER77}, see in particular p. 209). The four-velocity  tangent vector along $\mathcal {C}$  is $(\dot{t}, 0,0,\dot{t}d\phi/dt)$.
In a neighborhood of the circular orbit $\mathcal {C}$, all spacetime points may be charted using 
Fermi coordinates whose origin is the center of mass of the container of the gas.  Parker and Pimentel \cite{PP82} calculated the Fermi normal basis vectors, or tetrad, for a circular geodesic orbit at radius $r$  and gave some of the Riemann tensor components in that tetrad. Below, we give the Fermi basis vectors (after correcting a misprint in Eq. (5.24) of \cite{PP82}) and all of the nonvanishing Riemann tensor  components in the  Fermi frame.  Hereafter barred indices refer to the Schwarzschild coordinates of (\ref{a1}) while unbarred indices refer to Fermi coordinates $\{x^{0},x^{1},x^{2},x^{3}\}$.  As usual Greek indices run over $0,1,2,3$ and Latin over $1,2,3$.  Let,
\begin{equation}
\textbf{\textit{e}}_{\beta}=(e_{\beta}^{\;\bar{t}},e_{\beta}^{\;\bar{r}},e_{\beta}^{\;\bar{\theta}},
e_{\beta}^{\;\bar{\phi}}) \equiv (\frac{\partial t}{\partial x^{\beta}},\frac{\partial r}
{\partial x^{\beta}},\frac{\partial \theta}{\partial x^{\beta}},\frac{\partial \phi}{\partial
x^{\beta}}).
\label{a2}
\end{equation}

\noindent Then a parallel tetrad for Fermi coordinates is,

\begin{eqnarray}
\textbf{\textit{e}}_{0}&=&\left (\frac{\epsilon}{X},0,0,\frac{l}{r^{2}}\right ),\label{a3}\\
\textbf{\textit{e}}_{1}&=&\left (-\frac{l\sin(\alpha\phi)}{rX^{1/2}},X^{1/2}\cos(\alpha\phi),0,
-\frac{\epsilon\sin(\alpha\phi)}{rX^{1/2}}\right ),\label{a4}\\
\textbf{\textit{e}}_{2}&=&\left (0,0,\frac{1}{r},0\right ),\label{a5}\\
\textbf{\textit{e}}_{3}&=&\left (\frac{l\cos(\alpha\phi)}{rX^{1/2}},X^{1/2}\sin(\alpha\phi),0,
\frac{\epsilon\cos(\alpha\phi)}{rX^{1/2}}\right ),\label{a6}
\end{eqnarray}

\noindent where,
\begin{equation}
X=1-\frac{2M}{r},\;\;\;\;\;\;\;\;\;\;\;\;\;\alpha=\sqrt{\frac{r-3M}{r}},
\label{a7}
\end{equation}
$\epsilon$ is the energy per unit mass of a test particle following the orbit of the center of mass of the container of gas, and $l$ is its angular momentum per unit mass.  Thus,
\begin{equation}
\epsilon=\frac{r-2M}{\sqrt{r(r-3M)}},\;\;\;\;\;\;\mbox{and}\;\;\;\;\;\;\;l=r
\sqrt{\frac{M}{r-3M}}.
\label{a8}
\end{equation}
The tetrad vectors may be expressed explicitly in terms of the proper time
$x^{0}=\tau$ of the Fermi observer.  On the circular geodesic $\mathcal {C}$ we have that
\begin{equation}
\dot{\phi}=\frac{l}{r^{2}},
\label{i1}
\end{equation}
where the dot indicates the derivative with respect to the proper time along this geodesic,
namely $\tau$.  For the geodesic $\mathcal {C}$, $l$ is given by (\ref{a8}), therefore it follows
that
\begin{equation}
\phi=\frac{\sqrt{M}}{r\sqrt{r-3M}}\,\tau.
\label{i2}
\end{equation}
We now use the relation
\begin{equation}
R_{\alpha\beta\gamma\delta}=R_{\bar{\kappa}\bar{\lambda}\bar{\mu}\bar{\nu}}
e_{\alpha}^{\;\bar{\kappa}}e_{\beta}^{\;\bar{\lambda}}e_{\gamma}^{\;\bar{\mu}}
e_{\delta}^{\;\bar{\nu}},
\label{a9}
\end{equation}
to obtain the Riemann tensor components in the Fermi basis (cf. \cite{PP82}).

\begin{eqnarray}
R_{0101}&=&-\frac{M(r+3(r-2M)\cos (2\alpha \phi))}{2(r-3M)r^{3}},\label{a10}\\
R_{0103}&=&-\frac{3M(r-2M)\sin (2\alpha \phi)}{2(r-3M)r^{3}},\label{a11}\\
R_{0202}&=&\frac{M}{(r-3M)r^{2}},\label{a12}\\
R_{0303}&=&-\frac{M(r-3(r-2M)\cos (2\alpha \phi))}{2(r-3M)r^{3}},\label{a13}\\
R_{0113}&=&\frac{3M^{3/2}(r-2M)^{1/2}\cos (\alpha \phi)}{(r-3M)r^{3}},\label{a14}\\
R_{0212}&=&-\frac{3M^{3/2}(r-2M)^{1/2}\sin (\alpha \phi)}{(r-3M)r^{3}}.\label{a15}
\end{eqnarray}

\noindent From (\ref{a7}) and (\ref{i2}) we have that
\begin{equation}
\alpha \phi=\frac{\sqrt{M}}{r^{3/2}}\,\tau.
\label{i3}
\end{equation}
(\ref{i3}) expresses the argument of the trigonometric functions appearing in the the 
curvature components, (\ref{a10})-(\ref{a15}), in terms of $\tau$.\\
\indent In addition to the usual symmetries which connect the various components of the Riemann 
tensor we find the following relations
\begin{eqnarray}
R_{0313}&=&-R_{0212},\;\;\;\;R_{0223}=-R_{0113},\label{a16}\\
R_{1212}&=&-R_{0303},\;\;\;\;R_{1313}=-R_{0202},\label{aa16}\\
R_{1223}&=&-R_{0103},\;\;\;\;R_{2323}=-R_{0101}.\label{a17}
\end{eqnarray}

The metric components in Fermi coordinates to second order are  given (\cite{MM63}, \cite{Misner}, and \cite{P80}) by,

\begin{eqnarray}
g_{00}&=&-1-R_{0l0m}x^{l}x^{m}+O(x^{3}),\label{a18}\\
g_{0i}&=&g^{0i}=-\frac{2}{3}R_{0lim}x^{l}x^{m}+O(x^{3}),\label{a19}\\
g_{ij}&=&\delta_{ij}-\frac{1}{3}R_{iljm}x^{l}x^{m}+O(x^{3}),\label{a20}\\
g^{00}&=&-1+R^{0\;0}_{\;l\;m}x^{l}x^{m}+O(x^{3})\nonumber\\
&=&-1+R_{0l0m}x^{l}x^{m}+O(x^{3}),\label{a21}\\
g^{ij}&=&\delta^{ij}+\frac{1}{3}R^{i\;j}_{\;l\;m}x^{l}x^{m}+O(x^{3})\nonumber\\
&=&\delta^{ij}+\frac{1}{3}R_{iljm}x^{l}x^{m}+O(x^{3}),\label{a22}\\
g&=&-1+\frac{1}{3}(R_{lm}-2R_{0l0m})x^{l}x^{m}+O(x^{3}),\label{a23}
\end{eqnarray}
where $g=\det (g_{\alpha\beta})$.
 
\section*{{\normalsize 3 Phase space and the canonical partition function}}
The Hamiltonian for a system of $N$ ideal gas particles is given by
\begin{equation}
H=\sum_{I=1}^{N}H_{\scriptscriptstyle{I}},\;\;\;\;\mbox{where}\;\;\;\;H_{\scriptscriptstyle{I}}
=\frac{1}{2}g_{\scriptscriptstyle{I}}^{\alpha\beta}p_{\scriptscriptstyle{I}\scriptstyle{\alpha}}
p_{\scriptscriptstyle{I}\scriptstyle{\beta}},
\label{x27}
\end{equation}
Here $H_{\scriptscriptstyle{I}}$ is the Hamiltonian for the $I$th particle. For notational
simplicity we shall omit  the subscript $I$ labeling the particle coordinates and momenta, when it 
is clear that we are referring to a one-particle Hamiltonian.  The state of a single  particle
consists of its four spacetime coordinates together  with its four-momentum coordinates
$\{ x^{\alpha},p_{\beta}\}$. If the proper mass, $m >0$, of the  particle is fixed, then the
one-particle Hamiltonian is given by,
\begin{equation}
H_{\scriptscriptstyle{I}}=\frac{1}{2}g^{\alpha\beta}p_{\alpha}p_{\beta}=-\frac{m^{2}}{2}.
\label{a24}
\end{equation}
Eq. (\ref{a24}) allows us to reduce the dimension of the state space to seven. The 
energy of a test particle with four-momentum $p_{\beta}$ according to an observer with
four-velocity $u^{\alpha}$  in Schwarzschild spacetime, is $-u^{\alpha}p_{\alpha}$, which is a
scalar invariant under coordinate transformations. In the Fermi coordinates described in the
previous section, an observer with fixed space coordinates $\{x^{j}\}$ and time coordinate $\tau$,
e.g., the observer  at any fixed location in the container of the orbiting gas, with four-velocity
$u^{\alpha}= (1, 0, 0, 0)$, will measure energy  $-p_{0}$ of a particle with
four-momentum $p_{\beta}$ at that point.  With 
the assumption $p_{0} <0$ we obtain,

\begin{equation}
p_{0}=\frac{-g^{0i}p_{i}+\sqrt{(g^{0j}g^{0k}-g^{00}g^{jk})p_{j}p_{k}-g^{00}m^{2}}}{g^{00}}.
\label{a25}
\end{equation}

 Let $P$ be the sub bundle of the cotangent bundle of spacetime with each  three 
dimensional fiber determined by (\ref{a25}).\\

\noindent \textbf{Proposition 1}  \textit{The volume 7-form 
$\tilde{\omega}$ on $P$ given by, 
\begin{equation}
\tilde{\omega}=\frac{1}{p^{0}}dx^{0}\wedge dx^{1}\wedge dx^{2}\wedge dx^{3}\wedge dp_{1}\wedge dp_{2}\wedge dp_{3},\label{a26}
\end{equation}
is invariant  under all coordinate transformations.}\\

\noindent The proof of Proposition I which is analogous to a result  in \cite{C63} 
(see also \cite{Sachs}, p. 264 for generalizations) is given in Appendix A.

We next restrict our attention to the Fermi Observer along the circular geodesic 
$\mathcal{C}$ with fixed $\tau$.  Phase space for a single particle is determined by
the space slice orthogonal to the Fermi Observer's four-velocity, with the associated momentum
coordinates, along with the volume form given by the interior product (cf. \cite{Sachs}, p. 3) 
$\textbf{\textit{i}}(\partial/\partial\tau)\tilde{\omega}$ of the vector $\partial/\partial\tau$ 
with $\tilde{\omega}$. Then,

\begin{eqnarray}
m\,\textbf{\textit{i}}(\partial/\partial \tau)\tilde{\omega} &=& dx^{1}\wedge
dx^{2}\wedge dx^{3}\wedge  dp_{1}\wedge dp_{2}\wedge dp_{3}\nonumber\\
&\equiv& dx^{1} dx^{2} dx^{3} dp_{1}dp_{2}dp_{3}.
\label{new27}
\end{eqnarray}

\noindent It follows from Proposition 1
that the 6-form given by (\ref{new27}) is invariant under coordinate changes of the space 
variables (only), with $\tau$ fixed.  Physically this means that the
calculations that follow below are independent of the choice of the Fermi tetrad, (\ref{a2})-(\ref{a6}).  
We note in particular that with $\tau$ fixed, $p_{0}$ is invariant under
changes of the space coordinates.

Liouville's theorem for the invariance of phase space volume under the dynamics of 
particle motion is one of the foundations of statistical mechanics.  In order to 
establish a local version of Liouville's theorem associated with a product of $6$-forms each of 
the form given by (\ref{new27}) we return briefly to the cotangent bundle of spacetime and
consider the symplectic form, $\tilde{\omega}_{\,0}$, given by,

\begin{equation}
\tilde{\omega}_{\,0}=dx^{0}\wedge dp_{0}+dx^{1}\wedge dp_{1}+dx^{2}\wedge dp_{2}+dx^{3}
\wedge dp_{3}.\label{a27}
\end{equation}\

\noindent The Hamiltonian, $H_{\scriptscriptstyle{I}}$, governing the dynamics is given by (\ref{a24}). \textit{Henceforth, for the duration of this article, we use  $g_{\alpha \beta}$ to
represent the second order approximation to the Fermi metric given by (\ref{a18})-(\ref{a20})
rather than the exact (uncalculated) values.  Likewise, we use only the second order approximations
for the components of the inverse metric and $g^{\alpha \beta}$}. 
 
The metric and hence $H$ depend on $x^{0}$ only through the  Riemann tensor components. It follows from (\ref{a10})-(\ref{a15}) that the Hamiltonian is given to second order in all four spacetime 
coordinates $\{x^{0},x^{1},x^{2},x^{3}\}$ by setting $x^{0}=0$.  In fact, expanding to third order 
in the spacetime coordinates, it is not difficult to show that the time derivatives of the metric
tensor elements are much smaller than the derivatives in the spatial directions at spacetime points
close to the geodesic circular orbit (cf. \cite{MM63}, Section IX). For the purpose
of approximating statistical mechanical quantities we therefore make a further  approximation and 
set $x^{0}=0$ in (\ref{a10})-(\ref{a15}).  This amounts to selecting a  point on the circular
geodesic orbit and assuming an approximately time independent Hamiltonian. In light of the comments following Eq. (\ref{new27}), we would obtain the same final results from the selection of any other point on the orbit, and we briefly discuss this in Section 4 below. We therefore identify
$p_{0}=-E<0$, where the energy $E$ of a particle is positive and constant, and  it follows that we can reduce the
dimension of phase space by two by eliminating $p_{0}$ and its conjugate cyclic coordinate $x^{0}$.
  
We next consider an ideal gas consisting of $N$ particles in a one-dimensional box; the
generalization to additional degrees of freedom is straightforward and we omit it in order to keep
our indices simple.  We label  the coordinate, momentum, and proper time of the $I$th particle,
$(I=1,\dots,N)$ by $q_{\scriptscriptstyle{I}}, p_{\scriptscriptstyle{I}},
\tau_{\scriptscriptstyle{I}}$, respectively.  We assume that
$\tau_{\scriptscriptstyle{I}}=\tau_{\scriptscriptstyle{I}}(\tau)$, where $\tau$ the proper time  of
the center of mass of the system (in the interval between collisions).  Note that the function
$\tau_{\scriptscriptstyle{I}}(\tau)$ may be any smooth function of $\tau$.

We may define the Hamiltonian vector field, $\vec{v}_{H}$, as follows

\begin{equation}
\vec{v}_{H}={ \everymath{\displaystyle}
\left(
  \begin{array}{c}
     \dot{q}_{1} \\
     \rule{0in}{5ex}
     \dot{p}_{1} \\
     \rule{0in}{5ex}
     \vdots\\
     \rule{0in}{5ex}
     \dot{q}_{N} \\
     \rule{0in}{5ex}
     \dot{p}_{N}
  \end{array}
  \right)=
  \left(
  \begin{array}{c}
     \frac{dq_{1}}{d\tau} \\
     \rule{0in}{5ex}
     \frac{dp_{1}}{d\tau} \\
     \rule{0in}{5ex}
     \vdots\\
     \rule{0in}{5ex}
     \frac{dq_{N}}{d\tau} \\
     \rule{0in}{5ex}
     \frac{dp_{N}}{d\tau}
  \end{array}
  \right)=
\left(
  \begin{array}{c}
     \frac{dq_{1}}{d\tau_{1}}\frac{d\tau_{1}}{d\tau} \\
     \rule{0in}{5ex}
     \frac{dp_{1}}{d\tau_{1}}\frac{d\tau_{1}}{d\tau} \\
     \rule{0in}{5ex}
     \vdots\\
     \rule{0in}{5ex}
     \frac{dq_{N}}{d\tau_{N}}\frac{d\tau_{N}}{d\tau} \\
     \rule{0in}{5ex}
     \frac{dp_{N}}{d\tau_{N}}\frac{d\tau_{N}}{d\tau}
  \end{array}
  \right)}.\label{a28}
\end{equation}

\noindent Liouville's theorem, which states that the Hamiltonian system is measure-preserving, 
follows from (\ref{a29}):
\begin{eqnarray}
\mbox{div}\,\vec{v}_{H}&=&\frac{\partial \dot{q}_{1}}{\partial q_{1}}+\frac{\partial
\dot{p}_{1}}{\partial p_{1}}+\cdots+\frac{\partial \dot{q}_{N}}{\partial q_{N}}+\frac{\partial
\dot{p}_{N}}{\partial p_{N}},\nonumber\\
&=&\left[\frac{\partial}{\partial
q_{1}}\left(\frac{dq_{1}}{d\tau_{1}}\right)+\frac{\partial}{\partial
p_{1}}\left(\frac{dp_{1}}{d\tau_{1}}\right)\right]\frac{d\tau_{1}}{d\tau}+\cdots+
\left[\rule{0in}{3ex}\cdots\;\right]\frac{d\tau_{N}}{d\tau},\nonumber\\
&=&\left[\frac{\partial^{2}H}{\partial q_{1}\partial p_{1}}-\frac{\partial^{2}H}{\partial
p_{1}\partial q_{1}}\right]\frac{d\tau_{1}}{d\tau}+\cdots+
\left[\rule{0in}{3ex}\cdots\;\right]\frac{d\tau_{N}}{d\tau}=0.\label{a29}
\end{eqnarray}

\noindent We thus have a local version Liouville's theorem for our gas restricted to a small volume in 
circular orbit, that is, we may assert that Liouville's theorem holds to low order in the 
spacetime variables.

In order to proceed to the next step and define the thermodynamic free energy corresponding to 
$g_{\alpha \beta}$ (see (\ref{a32}) below) we will need the following theorem. The proof is given in Appendix B.\\

\noindent \textbf{Theorem 1}
\textit{There exists a positive function $f(M,r)$ defined for $r>3M>0$ such that:}\\

\noindent \textit{(a) $f(M,r)$ is a decreasing function of $M$ and an increasing function of 
$r$ and $f(M,r)\rightarrow\infty$ as $r\rightarrow\infty$ or $M\rightarrow 0$ when the other 
variable is held fixed.}\\ 

\noindent \textit{(b) If $|x^{i}|<f(M,r)$ for $i=1,2,3$, then}
\begin{equation}
p_{0}\leq -k(|p_{1}|+|p_{2}|+|p_{3}|),\label{a30} 
\end{equation}
\textit{for some positive number $k$, where $p_{0}$ is the function of space and momentum coordinates defined by (\ref{a25})}.\\

The canonical partition function for the gas under consideration depends on temperature. One 
would expect small variations in the temperature $T$ within the gas volume (a) because of
pressure variations due to gravity and (b) because of tidal effects on the particles in a
container in orbit around $M$ (cf. \cite{Tolman}, \cite{Klein}).  These effects, although very small,
would also be present in the nonrelativistic case.  We assume that our system is in contact with 
a `heat bath' and consequently in thermal equilibrium to a good approximation. Following the usual 
convention, we write,
\begin{equation}
\beta=\frac{1}{kT},
\label{a31}
\end{equation}
where $T$ is the temperature of the gas in volume $V$ and $k$ is Boltzmann's
constant.   Turning now to the canonical partition function, we write,

\begin{equation}
Z_{r}=\frac{1}{N!(2\pi\hbar)^{3N}} \int_{\mathbb{R}^{3N}}\int_{V^{N}}
e^{\,(\beta\sum_{I=1}^{N} (p_{0})_I}d\mathbf{x}^{N}d\mathbf{p}^{N},\label{a32}
\end{equation}

\[d\mathbf{x}=dx^{1}dx^{2}dx^{3}\,,\;d\mathbf{p}=dp_{1}dp_{2}dp_{3},\]
\begin{equation}
d\mathbf{x}^{N}d\mathbf{p}^{N}=dx^{1}_{1}dx^{2}_{1}dx^{3}_{1}dp_{11}dp_{12}dp_{13}\ldots
dx^{1}_{N}dx^{2}_{N}dx^{3}_{N}dp_{N1}dp_{N2}dp_{N3}.\label{a33}
\end{equation}
where $N$ is the number of particles, and $(p_{0})_{I}$ is given by (\ref{a25}) with the coordinates 
and momenta corresponding to the $I$th particle. The volume integrals in (\ref{a32}) have limits of integration determined by the range of Fermi coordinates that defines the volume of the gas.

The following result follows immediately from Theorem 1.\\

\noindent \textbf{Corollary 1}
\textit{The partition function given by (\ref{a32})
is finite for any particle number $N$, any inverse temperature $\beta$, and any volume 
$V$ for which $|x^{i}|<f(M,r)$  for $i=1,2,3$, where $f(M,r)$ is given by Theorem 1.}\\

\noindent \textit{Remark 1}
The physical significance of the bound  on $|x^{i}|$ given by 
$f(M,r)$, a function that increases with increasing $r$ or decreasing $M$, is that the Fermi coordinates of the volume of the container of the gas may be increased -- and therefore the volume itself may be enlarged -- as the orbital radius increases, or the mass of 
the central body decreases. Informally, in the limit as $r\rightarrow\infty$ or $M\rightarrow 0$, the volume of the gas may increase to infinity, and thus the thermodynamic limit may be considered with the bounds (\ref{a30}) on the integrand of the partition function (\ref{a32}).  We leave for future investigations the question of whether or not Theorem 1 can be sufficiently sharpened so as to permit consideration of thermodynamic limits for $M>0$ and finite radial orbits.\\

The connection with thermodynamics is given by the equations
\begin{equation}
F_{r}(\beta ,V,N)=-\frac{1}{\beta}\ln Z_{r},
\label{a34}
\end{equation}
where $F_{r}$ is the  Helmholtz free energy at radial coordinate $r$ for the orbit $\mathcal {C}$ 
of the gas for the metric $g_{\alpha \beta}$ defined by (\ref{a18})-(\ref{a23}).\\

\section*{{\normalsize 4 Special relativistic and Newtonian limits}}
From (\ref{a10})-(\ref{a17}) we see that the $R_{\alpha\beta\gamma\delta}$ vanish as
$r\rightarrow\infty$. Thus, from (\ref{a18}),(\ref{a19}), (\ref{a21}), (\ref{a22}), and 
(\ref{a25}),
\begin{equation}
g^{00}\rightarrow \eta^{00},\;\;\;\;g^{0i}\rightarrow 0,\;\;\;\;g^{ij}\rightarrow
\delta^{ij},
\label{a35}
\end{equation}
\begin{equation}
p_{0}\rightarrow -\sqrt{\delta^{jk}p_{j}p_{k}+m^{2}},
\label{a36}
\end{equation}
as expected.

Let $Z_{\infty}$ be the partition function as defined in (\ref{a32}) except with
$p_{0}$  replaced by its limiting value $-\sqrt{\delta^{jk}p_{j}p_{k}+m^{2}}$, as
$r\rightarrow\infty$.   The free energy for $Z_{\infty}$ is known to be the Helmholtz free energy 
$F_{\infty}(\beta ,V,N)$  for an ideal gas in special relativity.  An explicit calculation for
$F_{\infty}(\beta ,V,N)$ was  given, for example, in \cite{Pauli}.  In the next
theorem,  we show that when $r=\infty$,  so that the gas is far from the gravitational source,
$F_{r}(\beta ,V,N)$ becomes the free energy  of an ideal gas in special relativity.\\

\noindent \textbf{Theorem 2}
\textit{For any finite volume $V$,}
\begin{equation}
\lim_{r \rightarrow \infty} F_{r}(\beta ,V,N)=F_{\infty}(\beta ,V,N).
\label{a37}
\end{equation}

\noindent \textit{Proof} Since $f(M,r) \rightarrow \infty$ as $r \rightarrow \infty$, the 
condition $|x^{i}| <f(M,r)$ required by Theorem 1 is satisfied for all sufficiently large $r$. 
The result follows easily from the Lebesgue Dominated Convergence theorem for the interchange of 
limits and integrals because the integrand in (\ref{a32}) is dominated uniformly in
$r$ by an integrable function which is a product of $N$ terms of the form
$\exp(-k(|p_{1}|+|p_{2}|+|p_{3}|))\,.\;\;\square $\\

\noindent \textit{Remark 2} In Theorem 2, the Fermi coordinates for the volume $V$ relative to the orbital geodesic of radius $r$ are held constant, independent of $r$ (see Remark 1). Therefore, the volume of gas itself depends on $r$ in the sense that it is held in orbit at radius $r$, and its proper volume is a function of $r$, converging asymptotically to its special relativistic volume as $r \rightarrow \infty$. Alternatives are also possible. The proper volume of the gas container may be held fixed by choosing the limits of integration in the volume integrals in (\ref{a32}) to be suitable functions of $r$.\\

Curvature in relativistic spacetimes corresponds to tidal forces in Newtonian mechanics.  
For the purpose of identifying relativistic effects, we compute here the Newtonian canonical 
free energy and pressure of an ideal gas in a circular orbit around a massive body, subject 
only to tidal forces.

To that end, let $V$ be the volume of the gas orbiting a central body of mass $M$ in a circular 
orbit of radius $r$.  The origin of Cartesian coordinates is located at the center of the box with the $x^{1}$-axis along the radius of the orbit. Thus, at a particular instant in time, the coordinates of the center of mass of the central body  are $(r, 0,0)$. We note that it would be necessary to use a different coordinate system to prove Theorem 3 below, if we had chosen a point on the geodesic orbit, $\mathcal {C}$, other than that corresponding to $x^{0} = 0$ because of the dependence of the Newtonian limits of the Fermi tetrad given by Eqs. (\ref{a3}) - (\ref{a6}) on $\phi \propto x^{0}$.

Aside from elastic collisions with the wall of the container, the molecules, each of mass 
$m$, are in free fall subject only to tidal forces. The tidal force $F_{t}(x^{1},x^{2},
x^{3})$ can be expressed in terms of the gravitational force $F_{g}(x^{1},x^{2},x^{3})$.  
In our coordinate system,

\begin{eqnarray}
F_{t}(x^{1},x^{2},x^{3}) &=& F_{g}(x^{1},x^{2},x^{3}) - F_{g}(0,0,0),\nonumber\\\nonumber\\
&=& -GMm\left[\frac{(x^{1}-r,x^{2},x^{3})}{((x^{1}-r)^{2}+(x^{2})^{2}+(x^{3})^{2})^{3/2}} +
\frac{(r,0, 0)}{r^{2}}
\right],\label{a38}
\end{eqnarray}\\
\noindent The potential energy $\Phi$ satisfying $\Phi(0,0,0)=0$ and $F_{t}=-\nabla \Phi$ is 
given by\\

\begin{eqnarray}
\Phi (x^{1},x^{2},x^{3}) &=&\frac{-GMm}{\sqrt{(x^{1}-r)^{2}+(x^{2})^{2}+(x^{3})^{2}}}\nonumber\\
 &+&\frac{GMm(x^{1}+r)}{r^{2}}.
\label{a39}
\end{eqnarray}
For the purpose later on of comparing (\ref{a39}) with the potential energy obtained from the
nonrelativistic limit of our metric, we approximate (\ref{a39}) retaining terms to 
order 2 in the coordinates,
\begin{equation}
\Phi\approx -\frac{GMm(2(x^{1})^{2}-(x^{2})^{2}-(x^{3})^{2})}{2r^{3}}.
\label{a40}
\end{equation}

For a volume $V$ of an ideal gas of $N$ particles subject to an external field $\Phi$, 
the Hamiltonian is given by,

\begin{equation}
H = \sum_{I=1}^{N}\left(\frac{p_{\scriptscriptstyle{I}}^{2}}{2m}+\Phi
(\mathbf{x}_{\scriptscriptstyle{I}})\right),
\label{a45}
\end{equation}

\noindent where $\mathbf{x}_{\scriptscriptstyle{I}}$ denotes the position $(x,y,z)$ of the $I$th
particle.   The canonical partition function is then,
\begin{eqnarray}
Z(V,\beta, N)&=&\frac{1}{(2\pi\hbar)^{3N}N!} \int_{\mathbb{R}^{3N}}\int_{V^{N}}\exp(-\beta
H)d^{N}\mathbf{x}d^{N}\mathbf{p}, \nonumber\\
\rule{0in}{2ex}\nonumber\\
&=&\left(\frac{m}{2\pi\hbar^{2}\beta}\right)^{\frac{3N}{2}}\frac{1}{N!}\left(\int_{V}e^{-\beta
\Phi(x,y,z)}dx dy dz\right)^{N},\label{a46}
\end{eqnarray}
\noindent The Helmholtz free 
energy, $F(V,\beta , N)$, is given by,
\begin{equation}
-\beta F(V,\beta , N) = \ln Z(V,\beta, N)\,.\label{a47}
\end{equation}

\noindent We note that it follows that the equation of state is given by a local version of 
Boyle's Law \cite{Martin-Lof},
\begin{equation}
\beta P(\mathbf{x}) = \rho(\mathbf{x}),
\label{a48}
\end{equation}

\noindent where $\rho(\mathbf{x}) d\mathbf{x}$ is the probability density for finding a particle 
at position $\mathbf{x}$ given by,\\
 \begin{equation}
\rho (\mathbf{x})= \frac{N \exp (-\beta \Phi(\mathbf{x}))}{\int_{V}\exp (-\beta
\Phi(\mathbf{x})d\mathbf{x}}.
\label{a49}
\end{equation}

\noindent Integrating both sides of (\ref{a49}) over the volume gives $\langle P
\rangle V= NkT$, where 
$\langle P \rangle= \frac{1}{V}\int_{V}P(\mathbf{x})d\mathbf{x}$ is the averaged pressure.

We now consider the nonrelativistic limit of $p_{0}$, (\ref{a25}), which appears in 
the partition function (\ref{a32}).  We let
\begin{equation}
\lambda=\frac{1}{c},
\label{a41}
\end{equation}
and replace $M$ by $GM\lambda^{2}$ and $m$ by $m/\lambda$ in (\ref{a25}).  Since, in the
nonrelativistic limit, we discard the $mc^{2}$ term from the expression for the energy of each
particle (recall that $p_{0}c=-E$), it is convenient to define for the $I$th particle the 
function $h_{I}(\lambda)$ by

\begin{equation}
(p_{0}c+mc^{2})_{I}=\left(\frac{p_{0}}{\lambda}+\frac{m}{\lambda^{2}}\right)_{I}\equiv 
h_{I}(\lambda),
\label{b1}
\end{equation}
 and consider for the purpose of taking the nonrelativistic limit, instead of (\ref{a32}), the 
`reduced' partition function
\begin{equation}
\tilde{Z}_{\lambda}(V,\beta, N)=\frac{1}{N!(2\pi\hbar)^{3N}} \int_{\mathbb{R}^{3N}}\int_{V^{N}}
e^{\,\beta\sum_{I=1}^{N} h_{I}(\lambda)}d\mathbf{x}^{N}d\mathbf{p}^{N}.
\label{b2}
\end{equation}
\noindent The integrability of the integrand in (\ref{b2}) follows from the proof of Theorem 1
and Corollary 1. As before, The corresponding Helmholtz free  energy,  $\tilde F_{r}(V,\beta , N)$,
is given by,
\begin{equation}
-\beta \tilde F_{\lambda}(V,\beta , N) = \ln \tilde Z_{\lambda}(V,\beta, N).
\label{N2}
\end{equation}
The nonrelativistic limit of $h_{I}(\lambda)$ is given by
\begin{equation}
\lim_{\lambda\rightarrow 0^{+}}h_{I}(\lambda)=-\left(\frac{\textbf{\textit{p}}^{2}}{2m}-
\frac{GMm\left(2(x^{1})^{2}-(x^{2})^{2}-(x^{3})^{2}\right)}{2r^{3}}\right),
\label{a44}
\end{equation}
where $\textbf{\textit{p}}^{2}=p_{1}^{2}+p_{2}^{2}+p_{3}^{2}$.  Eq. (\ref{a44}) agrees with (\ref{a40}).

In what follows we let $G=1$ again.\\

\noindent \textbf{Theorem 3}
\textit{The Newtonian limit of the relativistic free energy given by (\ref{N2}) is the classical Helmholtz free energy, to second order in the space coordinates, for a gas subject to Newtonian tidal forces and otherwise ideal. Specifically, for any $\beta, N$, and $V$,
}\begin{equation}
\lim_{\lambda \rightarrow 0^{+}} \tilde F_{\lambda}(V,\beta , N)=F(V,\beta , N),
\label{N1}
\end{equation}
\textit{where the right hand side of (\ref{N1}) is given by (\ref{a47})}.\\

\noindent \textit{Proof}
In view of (\ref{b2}) and (\ref{a44}), it suffices to prove that,\\
\begin{equation}
\lim_{\lambda \rightarrow 0^{+}}\int _{\mathbb{R}^{3}}\int_{V}
e^{\,\beta h_{I}(\lambda)}d\mathbf{x}d\mathbf{p}=\int _{\mathbb{R}^{3}}\int_{V}
\lim_{\lambda \rightarrow 0^{+}}e^{\,\beta h_{I}(\lambda)}d\mathbf{x}d\mathbf{p},
\label{N3}
\end{equation}

\noindent for any index $I$, which for convenience of notation we omit for the duration of this 
proof.  Define $s(\lambda)\equiv \lambda p_{0}(\lambda)$. Factoring $\lambda$ into the square root
term in the equation for $p_{0}$, it is easy to see that a branch of the square root function may
be selected so that $s(\lambda)$ may be extended as an analytic function of $\lambda$ regarded as 
a complex variable within a sufficiently small disk centered at the origin of the complex plane
(depending on the mass term $m$).  We note in particular that the definition of
$s(\lambda)$ depends on the curvature tensor elements given by (\ref{a10})-(\ref{a15}). With 
$\phi \propto \tau=0$ these expressions all become rational functions of $\lambda$, except for 
(\ref{a14}) which upon replacing $M$ by $M\lambda^{2}$ is,
\begin{equation}
R_{0113}=\frac{3\lambda^{3}M^{3/2}(r-2M\lambda^{2})^{1/2}}{(r-3M\lambda^{2})r^{3}}.\label{x58}
\end{equation}
A branch of this function, analytic in a disk containing zero, exists.  Thus, all the nonvanishing
curvature elements have extensions to analytic functions of $\lambda$ in a neighborhood of the 
origin. We denote the analytic extension of $s(\lambda)$ again by
$s(\lambda)$.  The Taylor  expansion for $s$ centered at $\lambda = 0$ yields $s(0)=-m$ and
$\frac{\partial s}{\partial\lambda}(0)= 0$.  It follows immediately that the function
$\tilde{h}(\lambda) = (s(\lambda) + m)/\lambda^{2}$ is an analytic extension of
$h(\lambda)$ to the domain of $s(\lambda)$ with a removable singularity at $\lambda = 0$.  We
compute $\frac{\partial
\tilde{h}}{\partial \lambda}(0)= 0$ and, 

\begin{eqnarray}
12m^{3}r^{6} \frac{\partial^{2} \tilde{h}}{\partial \lambda^{2}}(0)
&=& 3r^{6}(p_{1}^{2} +p_{2}^{2}+p_{3}^{2})^{2}\nonumber\\
&+& 2Mm^{2}r^{3}p_{1}^{2}(6x_{1}^{2} - x_{2}^{2} - x_{3}^{2})\nonumber\\
&+& 2Mm^{2}r^{3}p_{2}^{2}(8x_{1}^{2} - 3x_{2}^{2} - 7x_{3}^{2})\nonumber\\
&+& 2Mm^{2}r^{3}p_{3}^{2}(8x_{1}^{2} - 7x_{2}^{2} - 3x_{3}^{2})\nonumber\\
&+& 2Mm^{2}r^{3}[8p_{2}p_{3}x_{2}x_{3}\nonumber\\
&-& 4p_{1}x_{1}(p_{2}x_{2}+p_{3}x_{3})]\nonumber\\
&-& 48m^{3}(M)^{3/2}r^{5/2}p_{1}x_{1}x_{3}\nonumber\\
&+& 48m^{3}(M)^{3/2}r^{5/2}p_{2}x_{2}x_{3}\nonumber\\
&+& 48m^{3}(M)^{3/2}r^{5/2}p_{3}(x_{1}^{2}-x_{2}^{2})\nonumber\\
&+& 36(Mr)^{2}m^{4}(x_{1}^{2}-x_{2}^{2}) \nonumber\\
&-& 9(M)^{2}m^{4}(x_{2}^{2}+x_{3}^{2}-2x_{1}^{2})^{2}\label{N4}.
\end{eqnarray} 

\noindent Restricting $\lambda$ once again to be real and positive, it now follows that
\begin{equation}
\lim_{\lambda\rightarrow 0^{+}} \frac{\partial h}{\partial \lambda}= 0,\label{aa1}
\end{equation}
and from (\ref{N4}) that
\begin{equation}
\lim_{\lambda\rightarrow 0^{+}}\frac{\partial^{2}h}{\partial \lambda^{2}}>0,\label{aa2}
\end{equation}
for any bounded volume $V\ni (x_{1}, x_{2},x_{3})$, and all $\textbf{\textit{p}}^{2}=p_{1}^{2}+
p_{2}^{2}+p_{3}^{2}> K$, for any sufficiently large positive constant $K$ (depending on $V$).  For
such $K$ define $A =\{(p_{1},p_{2},p_{3})\in \mathbb{R}^{3}: \textbf{\textit{p}}^{2} 
\leq K\}$ and denote the complement of $A$ in $\mathbb{R}^{3}$ by $A^{c}$. The left hand 
side of (\ref{N3}) may then be rewritten as,
\begin{equation}
\lim_{\lambda \rightarrow 0}\int _{A}\int_{V}
e^{\,\beta h(\lambda)}d\mathbf{x}d\mathbf{p} \,+ 
\lim_{\lambda \rightarrow 0}\int _{A^{c}}\int_{V}
e^{\,\beta h(\lambda)}d\mathbf{x}d\mathbf{p}.
\label{N5}
\end{equation}

\noindent The integrand in the first term of (\ref{N5}) is continuous in all of its variables
and  is therefore bounded on the compact region of integration and an interval for all $\lambda
\geqslant 0$ sufficiently small. From the Lebesgue dominated convergence theorem the limit may 
thus be interchanged with the integral.  In order to justify the interchange of the limit and
integrals for the second term of (\ref{N5}), we observe that $\frac{\partial^{2} h}{\partial
\lambda^{2}}(\lambda)>0$ for all $\lambda > 0$ sufficiently small, and thus $\frac{\partial
h}{\partial \lambda}(\lambda)> \frac{\partial h}{\partial \lambda}(0^{+})= 0$ for sufficiently 
small $\lambda >0$.  It follows that $h(\lambda)$ is an increasing function for $\lambda >0$
sufficiently small. The interchange of the limit and the integrals for the second term in (\ref{N5}) 
now follows from the integrability of $h(\lambda_{0})$ for fixed $\lambda_{0} >0$ and
the Lebesgue dominated convergence theorem. $\;\;\;\square $

\section*{{\normalsize 5 Conclusions}}
There is a dearth of examples in the scientific literature of calculations of specific statistical mechanical ensembles within the broader framework of general relativity. Even the simplest statistical ensemble -- the canonical ensemble for an ideal gas in a finite volume -- poses special technical problems in the context of general relativity, including even the existence of finite volume partition functions as finite integrals, addressed by the corollary to our Theorem 1.

Our investigation was motivated by a simple physical question: What
would be the gravitational effects on statistical mechanical experiments carried out and observed in
a spaceship in orbit around a large central mass? To test our results we found rigorous proofs of limiting cases: the nonrelativistic, Newtonian 
limit and the special relativistic limit for infinite radius of the circular orbit. 

Our methods apply, with minor changes, to an observer in circular orbit around a
Reissner-Nordstr\"{o}m charged mass.  Likewise using the results of \cite{BGJ05} one may consider the case of an observer in a general timelike equatorial orbit in the Kerr spacetime. A further generalization would be to include higher order terms in the spacetime variables in the expansion of the metric tensor in Fermi coordinates. A starting point for that purpose are the results in \cite{LN79}. 

The generalization to a gas of \textit{nonrelativistic} particles with interactions, for the purpose 
of studying the effects of the background curvature, is straightforward.  This can be
achieved by following the classical formulations and including a potential energy function of the
proper distance between particles in the Boltzmann factor in the partition function, (\ref{a32}).  For a more satisfactory treatment of interacting particle systems, it is possible that
the approaches taken in the context of special relativity, such as in \cite{HSP81} and \cite{MK81}, could be applied (with second order corrections) to the circumstances considered in this paper,  owing to the locally Minkowskian coordinates employed here.

Returning to the ideal gas case, in Section 8 of \cite{C64}, Chernikov found a covariant equilibrium 
single-particle Maxwell-Boltzmann distribution function expressed in Schwarzschild coordinates and constructed with help of an exact constant of the motion associated with a timelike Killing vector field.  That Killing field coincides with our $\{u^{\alpha}\}$ on the circular geodesic orbit $\mathcal {C}$.  However, there is no exact transformation law in closed form from Schwarzschild to the Fermi coordinates, only an approximate transformation law. So that approach necessarily yields approximate statistical mechanical ensembles, as in this paper.  It is also unclear how the Killing field may be exploited to prove an appropriate Liouville theorem in Fermi coordinates, which is the preferred coordinate system of the observer along the orbit, $\mathcal {C}$.\\

\noindent \textbf{\small Acknowledgements}
{\small The authors wish to thank Jacek Polewczak and Nicholas Kioussis for helpful discussions.}

\appendix
\section*{{\normalsize A Invariance of the 7-form.}}
\noindent \textit{Proof}
Consider a change of coordinates,
\begin{equation}
{x}^{\bar\alpha} = {x}^{\bar\alpha}(x^{0}, x^{1}, x^{2}, x^{3}),
\label{a56}
\end{equation}
and 
\begin{equation}
{p}_{\bar k} = \frac{\partial x^{\beta}}{\partial {x}^{\bar k}}\,p_{\beta}= \frac{\partial 
x^{0}}{\partial {x}^{\bar k}}\,p_{0}+\frac{\partial x^{i}}{\partial 
{x}^{\bar k}}\,p_{i}.
\label{a57}
\end{equation}

\noindent Denote the Jacobian by,

\begin{equation}
J\equiv J(x,p) \equiv \frac{\partial({x}^{\bar 0},{x}^{\bar 1},{x}^{\bar 2},{x}^{\bar 3},
{p}_{\bar 1},{p}_{\bar 2},{p}_{\bar 3})}{\partial(x^{0},x^{1},x^{2},x^{3},p_{1}, p_{2},p_{3})}.
\label{a62}
\end{equation}

\noindent Since the spacetime coordinates do not depend on momentum, it follows from a 
property of block triangular determinants that

\begin{equation}
J(x,p)=  \frac{\partial({x}^{\bar 0},{x}^{\bar 1},{x}^{\bar 2},{x}^{\bar 3})}{\partial
(x^{0},x^{1},x^{2},x^{3})}\,\,\frac{\partial({p}_{\bar 1},{p}_{\bar 2},{p}_{\bar 3})}
{\partial(p_{1},p_{2},p_{3})}\equiv J(x)\,J(p).
\label{a63}
\end{equation}

\noindent Substituting (\ref{a25}) for $p_{0}$ in the right side of (\ref{a57}), we find,

\begin{equation}
\frac{\partial {p}_{\bar k}}{\partial p_{i}} = \frac{\partial x^{i}}{\partial 
{x}^{\bar k}} -\frac{p^{i}}{p^{0}}\frac{\partial x^{0}}{\partial {x}^{\bar k}}.
\label{a61}
\end{equation}

\noindent Using (\ref{a61}) and the multilinearity of determinants, we may write,

\begin{equation}
J(p)=\frac{1}{p^{0}}\det\left( \begin{array}{cccc} 
p^{0} & p^{1} & p^{2} & p^{3} \\
\rule{0in}{5ex}
	\displaystyle{\partial x^{0}\over\partial {x}^{\bar 1}} & 
 \displaystyle{\partial x^{1}\over\partial {x}^{\bar 1}} & 
 \displaystyle{\partial x^{2}\over\partial {x}^{\bar 1}} & 
 \displaystyle{\partial x^{3}\over\partial {x}^{\bar 1}}\\
\rule{0in}{5ex}
	\displaystyle{\partial x^{0}\over\partial {x}^{\bar 2}} & 
 \displaystyle{\partial x^{1}\over\partial {x}^{\bar 2}} & 
 \displaystyle{\partial x^{2}\over\partial {x}^{\bar 2}} & 
 \displaystyle{\partial x^{3}\over\partial {x}^{\bar 2}}\\
\rule{0in}{5ex}
	\displaystyle{\partial x^{0}\over\partial {x}^{\bar 3}} & 
 \displaystyle{\partial x^{1}\over\partial {x}^{\bar 3}} & 
 \displaystyle{\partial x^{2}\over\partial {x}^{\bar 3}} & 
 \displaystyle{\partial x^{3}\over\partial {x}^{\bar 3}}\\
\end{array} \right).
\label{a64}
\end{equation}

\noindent Next, using the relationship, $p^{\alpha} = (\partial x^{\alpha}/ \partial 
{x}^{\bar\beta})\,{p}^{\bar\beta}$, we may rewrite the first row of the matrix in (\ref{a64}) so that,

\begin{equation}
J(p) = \frac{{p}^{\bar 0}}{p^{0}}\det \left[ {\partial x^{\alpha}\over \partial 
{x}^{\bar\beta}}\right].
\label{a65}
\end{equation}

\noindent Then substituting (\ref{a65}) into (\ref{a63}), we get,

\begin{equation}
J =  \frac{{p}^{\bar 0}}{p^{0}}\det \left[ {\partial {x}^{\bar\alpha}\over \partial 
x^{\beta}}\right] \det \left[ {\partial x^{\alpha}\over \partial 
{x}^{\bar\beta}}\right]=  \frac{{p}^{\bar 0}}{p^{0}},
\label{a66}
\end{equation}

\noindent proving the invariance of the 7-form.  $\;\;\;\square $

\section*{{\normalsize B Proof of Theorem 1.}}
\noindent The proof of Theorem 1 depends on three Lemmas.\\

\noindent \textbf{Lemma 1}
\textit{Let $B$ be a 3-by-3 real symmetric matrix with 
(necessarily real) eigenvalues $\lambda_{1} \leq \lambda_{2} \leq \lambda_{3}$.  Then for any 
column vector $\vec{p}\in \mathbb{R}^{3}$}
\begin{equation}
\vec{p}^{\;T}B\,\vec{p} \geq \lambda_{1} (p_{1}^{2}+p_{2}^{2}+p_{3}^{2}) \equiv \lambda_{1}
||\vec{p}\,||^{2},
\label{a50}
\end{equation}
\textit{where $T$ denotes transpose and $p_{i}$ is the $i$th component of $\vec{p}$.}\\

\noindent \textit{Proof}
Let $\mathcal{O}$ be the real unitary matrix that diagonalizes $B$, so that\\
\begin{eqnarray}
\vec{p}^{\;T}B\,\vec{p} &=& \vec{q}^{\;T}\mathcal{O}^{T}B\, \mathcal{O}\,\vec{q}\nonumber\\
&=&\vec{q}^{\;T}\left(\begin{array}{ccc}\lambda_{1} & 0 & 0 \\0 & \lambda_{2} & 0 \\0 & 0 & 
\lambda_{3}\end{array}\right) \vec{q}\nonumber\\
&=&\lambda_{1} q_{1}^{\,2}+\lambda_{2}q_{2}^{\,2}+\lambda_{3}q_{\,3}^{2}\nonumber\\
&\geq&\lambda_{1}||\vec{q}\,||^{2}\nonumber\\
&= &\lambda_{1}||\mathcal{O}^{T}\vec{p}\,||^{2}= \lambda_{1}
||\vec{p}\,||^{2},\nonumber
\end{eqnarray}
where $\vec{q} \equiv \mathcal{O}^{T}\vec{p}$ and where the last equation follows from 
the fact that unitary transformations preserve norms. $\;\;\;\square $\\

\noindent \textit{Remark 3}
The quadratic form $g^{jk}p_{j}p_{k}$ appearing in (\ref{a25}) 
may be expressed in the form $\vec{p}^{\;T}B\,\vec{p}$.\\

\noindent \textbf{Lemma 2}
\textit{For any $\epsilon>0$, there exists a positive function 
$f_{\epsilon}(M,r)$ defined for $r>3M>0$ such that:}\\

\noindent \textit{(a)  $f_{\epsilon}(M,r)$ is a decreasing function of $M$ and an increasing 
function of $r$.\\  
Moreover, $f_{\epsilon}(M,r)\rightarrow\infty$ as $r\rightarrow\infty$ or
$M\rightarrow 0$ when the  other variable is held fixed.}\\ 

\noindent \textit{(b) $|1+g^{00}|<\epsilon$, $|g^{0i}| < 2\epsilon/3$, $|g^{ij}|< \epsilon/3$ 
for $i\neq j$, and all eigenvalues of the 3-by-3 matrix whose $(i,j)$th component is
$g^{ij}$ exceed $1-\epsilon$  whenever $|x^{k}|<f_{\epsilon}(M,r)$ for $k=1,2,3$.}\\

\noindent \textit{Proof} Let $\epsilon>0$. From (\ref{a10})-(\ref{a17}), it follows that 
by choosing $|x^{k}|$ sufficiently small (for $k=1,2,3$) depending on $M$ and $r$,
$|R_{iljm}x^{l}x^{m} |< \epsilon$ for all $i,j$. Let  $f_{\epsilon}(M,r)$ be the supremum of such 
bounds on $|x^{k}|$.  The monotone and  limiting behavior of $f_{\epsilon}(M,r)$ follows from the
opposite monotone behavior of the Riemann  curvature tensor elements and their zero limits (see 
(\ref{a10})-(\ref{a17}) as $r\rightarrow \infty$ or $M\rightarrow 0$.  The bounds on the metric
tensor elements, inverse metric tensor  elements, (\ref{a19}), (\ref{a21}),
and (\ref{a22}) when $|x^{k}|<f_{\epsilon}(M,r)$.  The lower  bound for the eigenvalues follows
directly from the Ger\v{s}gorin Circle theorem for the location of  eigenvalues of a square 
matrix \cite{Horn}. $\;\;\;\square $\\

\noindent \textbf{Lemma 3}
\textit{For any positive real numbers $a, b$, and $c$,}
\begin{equation}
\sqrt{a^{2}+b^{2}+c^{2}}\geq \frac{1}{\sqrt{3}}(a+b+c).
\label{a51}
\end{equation}

\noindent \textit{Proof}
For any $x$ and $y$, $x^{2}+y^{2}\geq 2xy$. Thus,
\[a^{2}+b^{2}+c^{2} = \frac{1}{2}(a^{2}+b^{2})+\frac{1}{2}(a^{2}+c^{2})+\frac{1}{2}(b^{2}+c^{2})
\geq ab+ac+bc.\]
\noindent Multiplying both sides by $2/3$ and rearranging terms gives,
\[a^{2}+b^{2}+c^{2}\geq \frac{1}{3}(a+b+c)^{2}.\;\;\;\square\]

\noindent \textbf{Proof of Theorem 1}.
\noindent \textit{Proof}
From Lemmas 1, 2, and 3 with $|x^{k}|<f_{\epsilon}(M,r)$ 
for $k=1,2,3$,
\begin{eqnarray}
\sqrt{(g^{0i}p_{i})^{2}-g^{00}g^{jk}p_{j}p_{k}-g^{00}m^{2}}&>&\sqrt{-g^{00}g^{jk}p_{j}p_{k}}\,,
\nonumber\\
&\geq&\sqrt{(1-\epsilon)^{2}(p_{1}^{2}+p_{2}^{2}+p_{3}^{2})},\label{a52}\\
&\geq&\frac{1}{\sqrt{3}}(1-\epsilon)(|p_{1}|+|p_{2}|+|p_{3}|).\nonumber
\end{eqnarray}
\noindent Adding $-g^{0i}p_{i}$ to both sides and applying Lemma 2 yields,

\begin{eqnarray}
g^{00}p_{0}&>&-\frac{2\epsilon}{3}(|p_{1}|+|p_{2}|+|p_{3}|)\nonumber\\
&+&\frac{1}{\sqrt{3}}(1-\epsilon)(|p_{1}|+|p_{2}|+|p_{3}|).\label{a53}
\end{eqnarray}

\noindent Collecting terms, dividing by $g^{00}$, and again applying Lemma 2 yields,
\begin{equation}
p_{0}< -k(|p_{1}|+|p_{2}|+|p_{3}|),
\label{a54}
\end{equation}

\noindent for some positive number $k$ provided that $\epsilon<3/(3+2\sqrt{3})$. The function 
$f(M,r)$ may be defined to be $f_{\epsilon}(M,r)$ for any such value of $\epsilon$. $\;\;\;\square $

\end{document}